\documentclass[%
reprint,pra,
superscriptaddress,
amsmath,amssymb,
aps,
]{revtex4}

\usepackage{graphicx}
\usepackage{dcolumn}
\usepackage{bm}
\usepackage{hyperref}
\usepackage[utf8]{inputenc}
\usepackage{subfigure}
\usepackage{cleveref}
\usepackage{amsmath}

\begin{document}
	
\preprint{APS/123-QED}

\title{Imaging through scattering media via spatial-temporal encoded pattern illumination}
\author{Xingchen Zhao}%
\affiliation{%
	Texas A\&M University, College Station, Texas, 77843, USA}%
\author{Xiaoyu Nie}
\affiliation{%
	Texas A\&M University, College Station, Texas, 77843, USA}%
\affiliation{%
	Xi'an Jiaotong University, Xi'an, Shaanxi 710049, China}%
\author{Zhenhuan Yi}%
\affiliation{%
	Texas A\&M University, College Station, Texas, 77843, USA}%
\author{Tao Peng}%
\email{taopeng@tamu.edu}
\affiliation{%
	Texas A\&M University, College Station, Texas, 77843, USA}%
\author{Marlan O. Scully}%
\affiliation{%
	Texas A\&M University, College Station, Texas, 77843, USA}%
\affiliation{%
Baylor University, Waco, 76706, USA}%
\date{\today}

\begin{abstract}
Optical imaging through scattering media is a long-standing challenge. Although many approaches have been 
developed to focus light or image objects through scattering media, they are either invasive, restricted to stationary or 
slowly-moving media, or require high-resolution cameras and 
complex algorithms to retrieve the images. Here we introduce a computational imaging technique that can overcome these 
restrictions by exploiting spatial-temporal encoded patterns (STEP). We present non-invasive imaging through scattering media with a single-pixel photodetector. We show that the method is 
insensitive to the motions of media. We further demonstrate that our image reconstruction algorithm is much more efficient than correlation-based algorithms for single-pixel imaging, which may allow fast imaging in currently unreachable scenarios.
\end{abstract}

	\maketitle
	
\section{Introduction}
Optical imaging through turbid media has various applications from long range observations through turbulence to imaging inside living tissues \cite{lohmannSpeckleMaskingAstronomy1983, li2020single,li2021single,
ntziachristosGoingDeeperMicroscopy2010, yoonDeepOpticalImaging2020}. When light propagates through turbid media, the 
wavefront is distorted because of the inhomogeneity, 
resulting in the degradation of spatial resolution and the reduction of imaging depth. Low-order aberrations due to random fluctuations of refractive index can be overcome by adaptive optics \cite{tysonPrinciplesAdaptiveOptics2011} and turbulence-free ghost imaging \cite{meyersTurbulencefreeGhostImaging2011}, 
while the problem becomes intractable for optically opaque media in which strong light scattering scrambles the spatial information conveyed by light fields. Early experiments using holographic imaging~\cite{leithHolographicImageryDiffusing1966} demonstrated that scattering by stationary media does not erase spatial information carried by light fields~\cite{freundLookingWallsCorners1990}. Speckle patterns appear random but are essentially deterministic, and information about the optical input can be retrieved. Recent advances in wavefront shaping exploiting transmission 
matrix \cite{popoffMeasuringTransmissionMatrix2010, 
popoffImageTransmissionOpaque2010, yoonMeasuringOpticalTransmission2015, deaguiarEnhancedNonlinearImaging2016, kimTransmissionMatrixScattering2015} and optical phase conjugation \cite{heOpticalPhaseConjugation2002, 
yaqoobOpticalPhaseConjugation2008, wangFocusingDynamicTissue2015, hillmanDigitalOpticalPhase2013, vellekoopDigitalOpticalPhase2012} have realized focusing and imaging through scattering media. However, they are either invasive and require holographic or interferometric measurements or need prior knowledge of the scattering properties of the media. A more recent development taking advantage of the angular correlation of speckle pattern \cite{freundMemoryEffectsPropagation1988, osnabruggeGeneralizedOpticalMemory2017,liuPhysicalPictureOptical2019}, i.e., ``memory effect'', enables non-invasive imaging through scattering layers using the auto-correlation of speckle patterns and a phase-retrieval algorithm \cite{bertolottiNoninvasiveImagingOpaque2012, katzNoninvasiveSingleshotImaging2014, cuaImagingMovingTargets2017, liSingleshotMemoryeffectVideo2018}. Although these techniques do not depend on the scattering properties of the media, they share some common shortcomings such as the memory-effect range restricts this approach to thin scattering layers, the small single speckle grain requires high-resolution camera to resolve, and the iterative phase-retrieval algorithm suffers from falling into local optimal solution. 

In addition, the dynamic nature of some media, such as fog and biological tissues, introduces another aspect to the challenge: the mapping between input and output fields becomes time-dependent, resulting in a rapid decorrelation of optical output information. An interesting attempt to use speckle correlation and shower-curtain effect has realized the vision through the dynamic turbid medium between the object and camera~\cite{edreiOpticalImagingDynamic2016}. However, 
this method fails when the medium between the source and object is also dynamic. On the other hand, the dynamic 
nature can be leveraged for use via ultrasound-modulated light correlation to image an object hidden inside dynamic media ~\cite{ruanFluorescenceImagingDynamic2020a}. Nevertheless, the imaging speed is very slow due to the raster-scan-based measurements. Therefore, modalities that can provide fast imaging of objects completely immersed in dynamic media are still highly demanded. 

Here we report a computational imaging technique, termed spatial-temporal encoded pattern (STEP) illumination, that allows non-invasive imaging of an object through scattering media. We experimentally demonstrate that an image of the object can be reconstructed from a 1D time series of light intensity measured by a photodetector, with ground glass diffusers and slices of chicken breast (1.2 mm thick each) as the scattering media.

\section{Principle}
Enlightened by the concepts of intensity modulation and Fourier-transform-based discrimination~\cite{sudarsanamRealtimeImagingStrongly2016}, we design a sequence of patterns that consists of a bundle of sinusoidal time series with different frequencies, such that every spatial location in each pattern is encoded by a unique frequency which is also a unique feature of the periodic oscillation of the pixel values along ``time'' axis (\textit{i.e.}, looking at a single spatial location through different patterns). We illuminate the diffuser-object system with this sequence of patterns and collect the transmitted light by a single-pixel detector. With the help of an image reconstruction algorithm based on fast Fourier transform (FFT), the images of the objects can be retrieved without prior knowledge of the objects and the scattering media.
\begin{figure}[hbtp]
    \centering
    \includegraphics[width=0.85\linewidth]{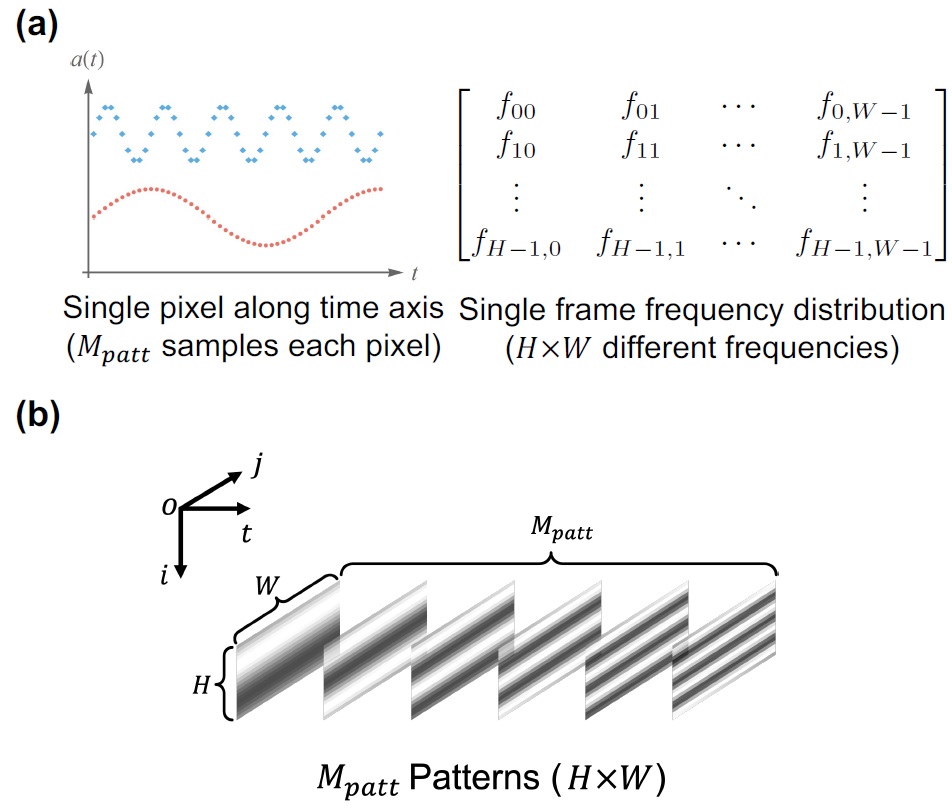}
    \caption{The design of STEP. (a). The sequence of patterns consists of a bundle of sinusoidal time series with unique frequency at each spatial location. (b). Grayscale pattern sequence used to image an object through scattering media.}
    \label{fig:fig1_principle}
\end{figure}

The design of STEP is sketched in Fig.~\ref{fig:fig1_principle}. We generate a 
sequence of grayscale (8-bit, 256 pixel values) patterns of height $H$ and width $W$ ($H\times W$ matrix). For a spatial location $\left(i,j\right)$ ($i$th row, $j$th column), the time series is given by
\begin{equation}\label{eq:patt_sigal_def}
	a_{ijt}=127.5 \sin{2\pi f_{ij} \frac{t}{r_s}}+127.5, 
	t=0,1,2,\cdots M_{\mathrm{patt}}-1,
\end{equation}
in which $r_s$ is the sample rate, $t$ is the discrete time variable (index of the patterns), $M_{\mathrm{patt}}$ is the total number of patterns. The frequency $f_{ij}$ is defined by $f_{ij}=f_{0}+\left(j+iW\right)\Delta f$ where $i=0,1,\cdots,H-1, j=0,1,\cdots,W-1$, are the row and column index of the pixels, respectively. $f_0$ is the starting frequency, and $\Delta f$ is the increment. To avoid signal aliasing, we set $r_s=8 f_{\mathrm{max}}$, where 
$f_{\mathrm{max}}=f_0+(HW-1)\Delta f$. The patterns are successively projected to the diffuser-object system and synchronically collected by the single-pixel photodetector. The 1D time series intensity can be expressed by
\begin{equation}\label{eq:1D_ts_intensity}
	I_t=\sum_{i,j}{a_{ijt}I_{ij}}+N_t,  
\end{equation}
where $I_{ij}$ is the illumination intensity at location $\left(i,j\right)$ of the pattern displayed on DMD and $N_t$ is a white noise term describing the noise of detector and environment. To reconstruct an image, $I_{t}$ is transformed to the spectral domain
\begin{equation}\label{eq:FT_discrete_I}
	S\left(\omega\right)=\mathcal{F}_{t}\left\{ I_{t}\right\} \propto M_{\mathrm{patt}}\sum_{i,j}I_{ij}\delta\left(f-f_{ij}\right)+\bar{N}
\end{equation}
where $\mathcal{F}_{v}$ denotes Fourier transform with respect to the variable $v$, $M_{\mathrm{patt}}$ happens to be the number of data points in the discrete 1D time series (integration length) since the measurement is synchronized with the projection, and 
$\bar{N}$ is the average noise level. For each $f_{ij}$ in the patterns, we find the closest frequency $\hat{f}_{ij}$ in the spectrum and save its magnitude $S(\hat{f}_{ij})$. Finally, a $H\times W$ matrix is filled with all the $S(\hat{f}_{ij})$ in their locations $\left(i,j\right)$, and a heat map of this matrix will yield an image of the object. It is the one-to-one correspondence between the frequency $f_{ij}$ and the spatial location $\left(i,j\right)$ that allows us to retrieve the spatial information computationally, and therefore we only need to measure the transmitted light with a single-pixel detector. We note here that the noise only contributes a constant term in the spectrum given by Eq. (\ref{eq:FT_discrete_I}).

\section{Results}
\begin{figure}[!ht]
    \centering
    \includegraphics[width=0.95\linewidth]{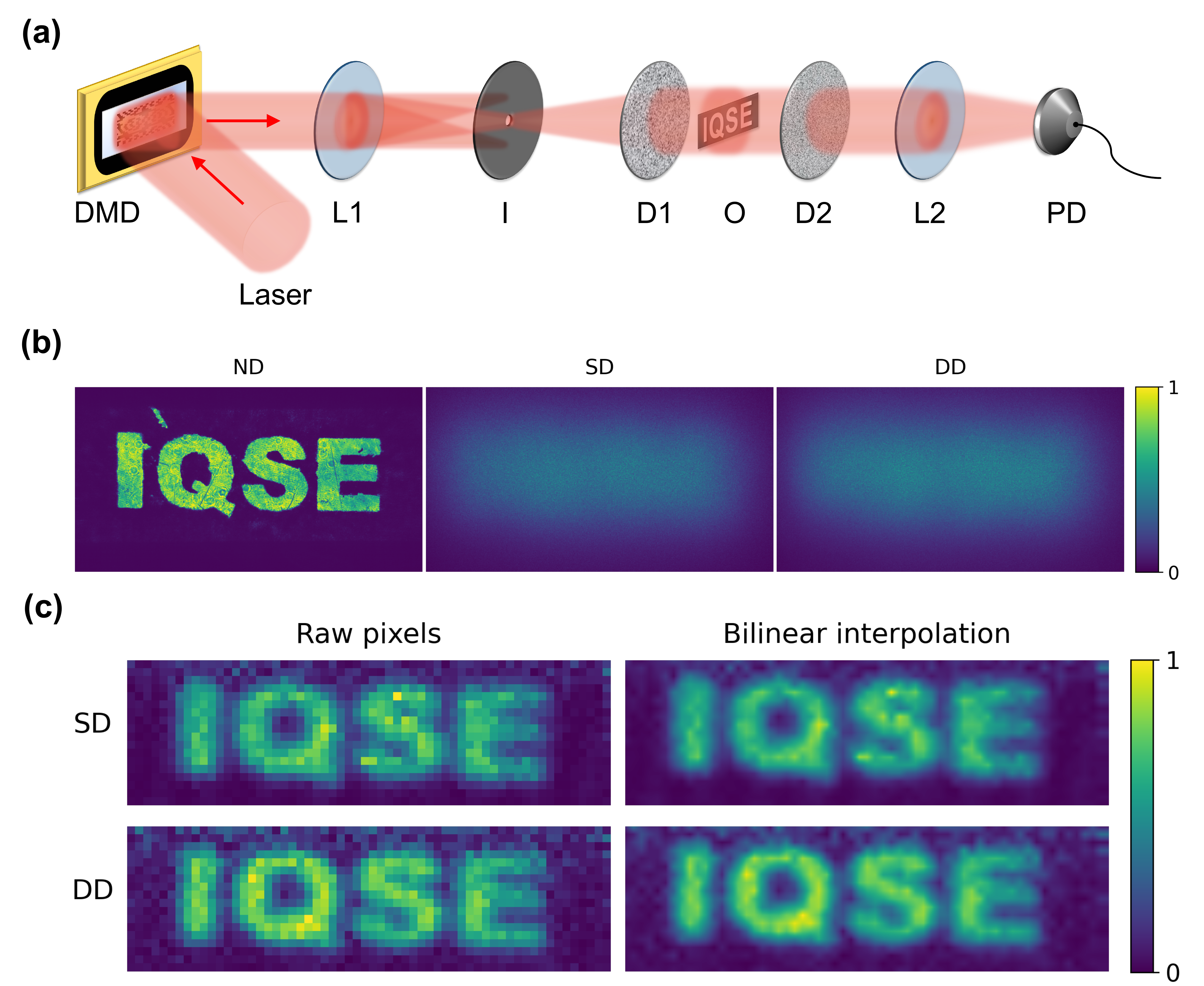}
    \caption{Experimental demonstration of STEP with ground glass diffusers. (a) Schematic of the setup. DMD: digital micromirror device; L1 and L2: lenses; I: iris; D1 and D2: diffusers; O: object; PD: photodetector. (b) Images captured by a CMOS camera under three conditions: without scattering media (ND), with stationary diffusers (SD), and with dynamic diffusers (DD). (c) Images reconstructed by STEP with $\beta=8$. Bilinear interpolation is applied to remove the pixelation effect.}
    \label{fig:fig2_GGD}
\end{figure}

A schematics of the experiment setup is shown in Fig. \ref{fig:fig2_GGD}(a). The grayscale patterns are first decomposed to $8M_{\mathrm{patt}}$ monochrome (1-bit, 2 pixel values) patterns in order to be compatible with the input format of the DMD. A solid-state laser (633 nm) is used to illuminate the DMD, which spatially modulated the incident laser beam and generate a set of illumination projections with a spatial structure that is similar to the input monochrome patterns. Each projection has $20\times 60$ pixels, and each pixel is maintained by a $10\times 10$ array of DMD micromirrors (each micromirror has a size of $10.8\ \mu m \times 10.8\ \mu m$). We define such an array of mirrors as a single ``DMD pixel''. The patterns are then imaged by a lens (L1) with a magnification of 2. High-order images due to diffraction are filtered out by an iris (I) such that only the zeroth-order image with the strongest intensity is formed in the image plane. A high-contrast object (O) with letters ``IQSE'' ($3\ \text{mm}\times 10\ \text{mm}$) being transparent is placed at the image plane of L1, where the zeroth-order images of the patterns are directly projected on the region of ``IQSE'' without scattering media. Then, two ground glass diffusers (D1 and D2, 220 grit) are inserted to block the view of the object. The distances between the object and D1 and D2 are about 1 mm and 5 mm, respectively. The diffusers can be kept stationary or moved back and forth by a motorized stage. A lens (L2) followed by a photodetector (PD) is placed behind D2 to collect the transmitted light.

The image of the object is first captured by a CMOS camera (replace the PD) under three conditions: without scattering media (ND), with stationary diffusers (SD), and with dynamic diffusers (DD). As shown in Fig. \ref{fig:fig2_GGD}(b), the object is invisible in the camera images  when the diffusers are present (SD and DD); while the STEP imaging scheme can retrieve the images through both SD and DD (Fig.~\ref{fig:fig2_GGD}(c)). We define $\beta=M_{\mathrm{patt}}/M_{\mathrm{pixel}}$
as a scaled number of patterns, in which $M_{\mathrm{patt}}$ is the number of grayscale patterns and 
$M_{\mathrm{pixel}}=120$ is the number of pixels in one pattern. The measurements in Fig.~\ref{fig:fig2_GGD} are performed with $\beta=8$, \textit{i.e.}, 9600 gray-scale patterns 
($8\times 96000$ monochrome patterns). To reconstruct an image, the frequency resolution of the Fourier transform $\delta f = r_{s}/M_{\mathrm{patt}}$ must satisfy the condition $\delta f\geq \Delta f$, which determines the minimum value of $\beta$. Our experimental parameters ($f_0=\Delta f=0.1$, $H=20$, $W=60$) give $\beta\geq 8$. Images in Fig. \ref{fig:fig2_GGD}(b) are obtained with $\beta=8$. Raw pixels in the reconstructed image are always the same as those in the patterns. Such a small number of pixels leads to pixelated images. Nevertheless, the pixelation effect can be eliminated computationally by applying bilinear interpolation to the raw pixels, and interpolated images of size 
$400\times 1200$ are given in Fig. \ref{fig:fig2_GGD}(b).

The size of an individual pixel determines the imaging resolution in the patterns projected onto the object: the smaller the pixel is, the more details of the object can be resolved. The pixel size at the image plane of L1 is determined by the size of the DMD pixel and the magnification of the imaging system defined by L1 when there are no scattering media. Decreasing the DMD pixel size will increase the resolution for a given magnification. However, the optical power reflected by each DMD pixel will be reduced due to the shrinkage of the reflective area (fewer micromirrors), resulting in a decreased signal-to-noise ratio (SNR) of the measured light. Conversely, increasing the DMD pixel will provide better SNR at the expense of a lower resolution. If the scattering layers are inserted, the degradation of pattern quality and the attenuation of optical power due to scattering must also be considered. Therefore, the trade-off between resolution and SNR should be decided according to the configuration of a specific setup. In our experiment, the resolution is $\sim 0.2$ mm, which is adequate for resolving the object of size $3\ \text{mm} \times 10\ \text{mm}$,, and the transmitted light intensity is far above the shot noise level of the detector.

The imaging depth is determined by how well the structure of the patterns can be preserved after transmitting through the first scattering layer and before illuminating the object. This is related to the so-called shower-curtain effect: an object placed at a distance behind a scattering layer looks blurred, but if the object is attached to the layer, it can be clearly seen \cite{edreiOpticalImagingDynamic2016}. Within a short distance after D1, the relative positions between the pixels on the patterns may be unchanged even after scattering. The one-to-one correspondence between the spatial location and the frequency of the sinusoidal intensity is retained. Since the measured signal is a convolution of the patterns and the object, the spatial information can be retrieved by extracting the frequencies $f_{ij}$, and an image of the object can be produced by $S(f_{ij})$. The shower-curtain effect indicates that the imaging depth of STEP is shallow ($\lesssim 1$ mm for current setup). Despite this, STEP is inherently insensitive to the scattering properties of D2 and its distance from the object because any scattering event cannot destroy the spatial information carried by the convoluted signal. In addition, The imaging speed of STEP is only limited by the frame rate of the DMD, as modern computers can complete the computation in little time.
The highest frame rate of current commercial DMD is 22.7 kHz, which will reduce our measurement time to a few seconds for $\beta=8$. 
\begin{figure}[!ht]
    \centering
    \includegraphics[width=0.95\linewidth]{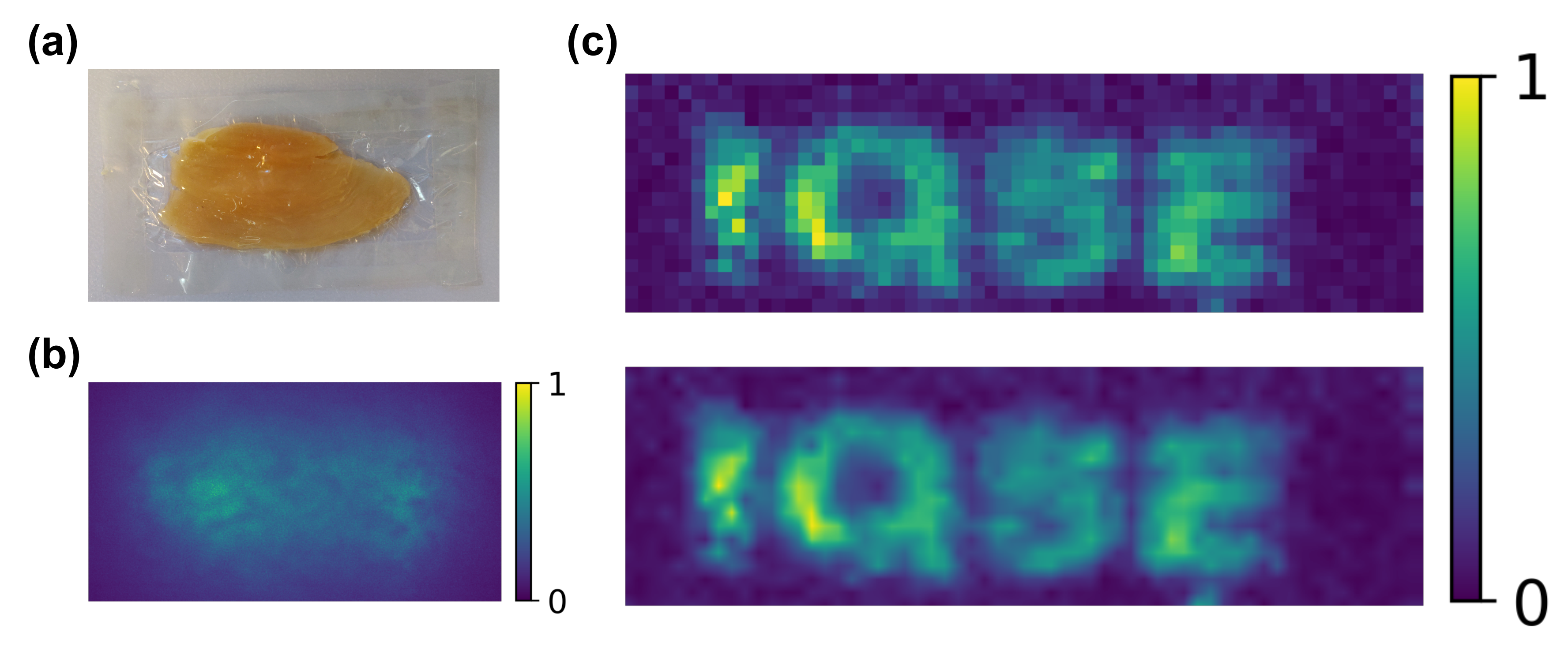}
    \caption{Imaging through two slices of chicken breast ($\sim 1.2$ mm each slice) with STEP. (a) One of the chicken breast slices used in the experiment, which is sealed in plastic wraps. (b) A camera image of the object hidden between two chicken breast slices. (b) Image reconstructed by STEP with $\beta=96$ (top). Bilinear interpolation is applied to remove the pixelation effect (bottom).}
    \label{fig:fig3_CB}
\end{figure}

To further demonstrate that STEP is also insensitive to the motion of the scattering centers in the media, we replaced the ground glass diffusers in Fig. \ref{fig:fig2_GGD}(a) with two slices of chicken breast ($\sim 1.2$ mm) and performed similar measurements with $\beta=96$. As shown in Fig.~\ref{fig:fig3_CB}(b), the object cannot be resolved in the camera image, whereas the image can be reconstructed by STEP (Fig. \ref{fig:fig3_CB}(c)). The larger $\beta$ is, the more capable the spectrum can distinguish target frequencies from the noisy background. A much larger $\beta$ to perform the computation implies a stronger tissue scattering than in the ground glass diffusers. Due to the nonuniform texture of the tissue, the transmitted light intensity is not as evenly distributed as that for the GGDs, resulting in some bright spots in the reconstructed image. Nevertheless, the results support the argument that STEP can image through dynamic scattering media.
\begin{figure*}[!htbp]
    \centering
    \includegraphics[width=0.95\linewidth]{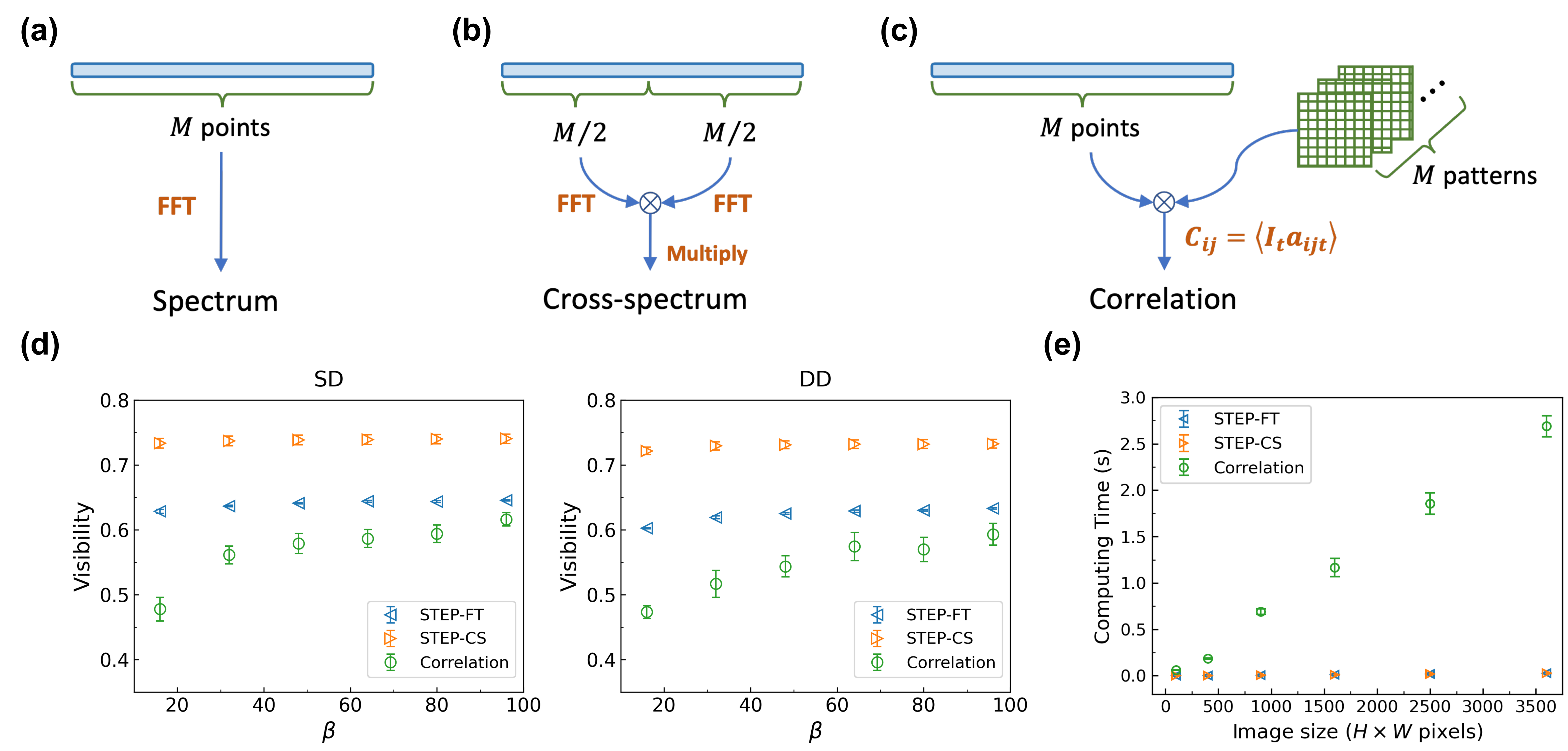}
    \caption{Comparison of different image reconstruction algorithms. (a) Fourier transform (FT): a segment of $M$ data points in the time domain is transformed by FFT to frequency domain (spectrum). (b) Cross-spectrum (CS): a segment of $M$ data points are divided into two halves, and their CS is calculated. (c) Correlation: the correlation between a segment of $M$ data points and the time series in the original pattern sequence is calculated. (d) Visibility of the reconstructed images using the three algorithms with different values of $\beta$. (e) Comparison of time complexity of the image reconstruction algorithms. Computing time is measured with varying sizes of images in total pixels.}
    \label{fig:fig4_vis_time}
\end{figure*}

We define the visibility of the reconstructed image to be $v=\left(\bar{p}_{s}-\bar{p}_{b}\right)/\left(\bar{p}_{s}+\bar{p}_{b}\right)$
in which $\bar{p}_s$ and $\bar{p}_b$ are the average pixel values of the signal (``IQSE'' regions) and background (other regions), respectively. Different from conventional structured illumination techniques in which the information required for image reconstruction is only encoded in the spatial structure of the patterns \cite{sun3DComputationalImaging2013, duranCompressiveImagingScattering2015, zhangSinglepixelImagingMeans2015, huSinglepixelPhaseImaging2019, gibsonSinglepixelImaging122020}, the STEP illumination also encodes the information (frequencies) in time. One advantage of introducing the time-domain encoding is that it allows a further improvement of the visibility via halving a successive segment of the data and calculating the cross-spectrum of the two halves. The cross-correlation technique  has been demonstrated very effective in weak signal detection \cite{liDetectingTechniqueWeak2003, mohideenabdulrazakDetectionExtractionWeak2009, rudnickDetectionWeakSignals1953}. Cross-spectrum is the frequency-domain representation of the cross-correlation of two time series signals, which is defined by $S_{12}\left(f\right)\equiv\mathcal{F}_{\tau}\left\{ g_{1}\star g_{2}\right\} =\mathcal{F}_{t}\left\{ g_{1}\right\} \cdot\mathcal{F}_{t}\left\{ g_{2}\right\}$, where $\tau$ is the delay in time, and $g_{1}\star g_{2}=\sum_{t=0}^{M}g_{1,t}g_{2,t+\tau}$ is the cross-correlation of two discrete signals $g_{1}$ and $g_{2}$ with $M$ data points. For images reconstructed with a specific $\beta$, higher visibility is obtained if the searching of target frequencies is performed in the cross-spectrum of the two halves of the data (Fig.
~\ref{fig:fig4_vis_time}(b)) rather than in the Fourier spectrum of all the data (Fig.~\ref{fig:fig4_vis_time}(a)). As shown in Fig.~\ref{fig:fig4_vis_time}(b), we divide the whole data set into two segments of the same length, and their cross-correlation is found to be  
\begin{equation}\label{eq:ccorr_discrete_I}
	I_{1}\star I_{2}=\frac{M}{2}I_{ij}I_{ij}^{\prime}\sin{2\pi f_{ij}\frac{\tau}{r_{s}}}+\frac{M}{2}\bar{N}_{1}\bar{N}_{2}
\end{equation}
where $M$ is the total length of the two signal segments, and $\bar{N}_{1}$ and $\bar{N}_{2}$ are the average noise levels for the two segments, respectively. It follows that the cross-spectrum is
\begin{equation}\label{eq:cs_discrete_I}
	S_{12}\left(f\right)=\mathcal{F}_{\tau}\left\{ I_{1}\star I_{2}\right\} 
	\propto\frac{M}{2}\sum_{ij}I_{ij}I_{ij}^{\prime}\delta\left(f-f_{ij}\right)+
	\frac{M}{2}\delta\left(0\right)
\end{equation}
As we have seen in Eq.~(\ref{eq:FT_discrete_I}), the noise contributes a constant background equally at every frequency in the spectrum. However, in the cross-spectrum (Eq.~(\ref{eq:cs_discrete_I})), the noise is concentrated to zero frequency (indicated by $\delta\left(0\right)$) and never contributes to the cross-spectral magnitude at any other frequencies. Therefore, the SNR is enhanced, leading to sharper peaks of the target frequencies and less noisy reconstructed images. 

The image reconstruction can also be implemented by calculating the correlations between the measured intensity data $I_{t}$ and the time series $a_{ijt}$ in the original patterns $C_{ij}=\sum^{M-1}_{t=0}I_t a_{ijt}$ \cite{sun3DComputationalImaging2013, duranCompressiveImagingScattering2015, gibsonSinglepixelImaging122020}. Filling an $H \times W$ matrix with all the $C_{ij}$ at their locations $\left(i,j\right)$ will yield an image of the object. However, the visibility of the image reconstructed by correlation is lower than that of the image produced by FFT-based methods, 
because in this case the correlation is done without a shift in time, leading to a non-trivial contribution of the white noise. Fig. \ref{fig:fig4_vis_time}(d) compares the visibility of the images generated by the three algorithms for different $\beta$. It is worth mentioning that the performances of STEP-FT and STEP-CS are almost independent of $\beta$: they have similar visibility over the investigated range of $\beta$ and show  saturation behaviors. This means high-quality images may be obtained with small data sets, and thus consume less time on measurement and computation. On the other hand, the correlation has the worst overall performance and is sensitive to the value of $\beta$. Therefore, high-quality images may only be acquired with a large number of data points.

The correlation method also suffers from a low computational efficiency. For $N$ patterns of size $H\times W$, both the time and space complexity of the image reconstruction via correlation are $O(HWN)$, in terms of the big-O notation; whereas, the time complexity of FFT-based reconstruction (FT and CS) is $O\left(N\mathrm{log}_{2}N\right)$ since the FFT algorithm is used to compute the spectrum, and the space complexity is $O\left(N\right)$ as there is no need to store the original patterns. A benchmark of the computing time for FFT-based and correlation algorithms are given in Fig. \ref{fig:fig4_vis_time}(e), where the rapid increase in the computing time for correlation algorithm provides a sharp contrast with 
those of the FFT-based methods. The high computational efficiency of the FFT-based algorithm may enable fast image processing with devices having limited computing resources.

\section{conclusions}
In conclusion, we have developed a computational imaging method named ``STEP'' that can realize non-invasive imaging through scattering media with a single-pixel photodetector. This method is insensitive to the motion of the scattering centers in the media. The design of STEP removes the requirement of a high-resolution camera and allows an elegant FFT-based image reconstruction algorithm that is more computationally efficient than correlation-based methods, which may be more favorable in many application fields. Our technique provides a new perspective to realize the vision of peeking through turbid media and enables potential fast imaging for currently unreachable scenarios. 

\section{Funding.}
Air Force Office of Scientific Research (Award No. FA9550-20-1-0366 DEF), Office of Naval Research (Award No. N00014-20-1-2184), Robert A. Welch Foundation (Grant No. A-1261), National Science Foundation (Grant No. PHY-2013771).


\end{document}